\documentclass{LT23auth}
\usepackage{graphicx}

\begin{document}

\begin{frontmatter}

\title{ESR Study of (C$_5$H$_{12}$N)$_2$CuBr$_4$}

\author[address1]{S. Zvyagin},
\author[address2]{B. C. Watson},
\author[address2]{Ju-Hyun Park}
\author[address3]{D. A. Jensen},
\author[address3]{A. Angerhofer},
\author[address1]{L.-C. Brunel},
\author[address3]{D. R. Talham},
\author[address2]{M. W. Meisel\thanksref{thank1}}

\address[address1]{National High Magnetic Field Laboratory, Florida State 
University, Tallahassee, FL  32310, USA.}
\address[address2]{Department of Physics and National High Magnetic Field 
Laboratory, University of Florida, Gainesville, FL  32611-8440, USA.}
\address[address3]{Department of Chemistry, University of Florida, Gainesville, 
FL  32611-7200, USA.}

\thanks[thank1]{Corresponding author. E-mail: meisel@phys.ufl.edu}

\begin{abstract}
ESR studies at 9.27, 95.4, and 289.7 GHz have been performed on 
(C$_5$H$_{12}$N)$_2$CuBr$_4$ 
down to 3.7 K.  The 9.27 GHz data were acquired with a single crystal and do 
not indicate the presence of any structural transitions.  The high frequency data 
were collected with a polycrystalline sample and resolved two 
absorbances, consistent with two crystallographic orientations of the 
magnetic sites and with earlier ESR studies 
performed at 300 K.  Below $B_{C1}=6.6$ T, our data 
confirm the presence of a spin singlet ground state.
\end{abstract}

\end{frontmatter}

Recently, a considerable amount of attention has been focused on the 
properties of quantum spin liquids \cite{Dagotto,Wessel}.
Herein, we report the results of ESR studies of 
(C$_{5}$H$_{12}$N)$_{2}$CuBr$_{4}$,
hereafter BPCB, 
which has been identified as a magnetic $S = 1/2$ two-leg  ladder 
in the strong coupling limit \cite{Watson}.  The crystal
structure of BPCB has been determined at 300 K to be monoclinic, 
and the Cu$^{2+}$ ions are stacked
in pairs that form a ladder-like structure \cite{Patyal}.
The magnetization possesses two critical fields, 
$B_{C1} = 6.6$ T and $B_{C2} = 14.6$ T, that were calculated from  
exchange interactions along the legs ($J_{\parallel} = 3.8$ K) and along 
the rungs ($J_{\bot} = 13.3$ K).  
Our motivation was to investigate the excitation spectrum of BPCB in 
different regions of its phase 
diagram.

The 9.27 GHz ESR work utilized standard X-band techniques \cite{Watson}.   
The high frequency measurements used a transmission-probe equipped with a 
millimeter and sub-millimeter wavelength band spectrometer with oversized
waveguides \cite{Brunel} and operated in a Faraday configuration down to 3.7 K.

For the 9.27 GHz work, a powder sample and a 18.6 mg single crystal specimen were 
measured at 300 K and between $20-300$ K, respectively.  For the single crystal, 
$B$ was oriented along the c-axis.  Two different 
samples were used for the high frequency studies, and both yielded similar results.  
Each of the samples consisted of a polycrystalline plate-like piece.  
Neutron studies on similar pieces showed that the constituent crystallites 
are generally oriented with the b-axis perpendicular to the plane of the larger 
plate.  For the high frequency studies, $B$ was oriented perpendicular to the 
plates, or generally parallel to the crystallographic b-axis.  
All of the samples were taken from the same batch as used previously \cite{Watson}.


At 9.27 GHz, the 300 K results were consistent with the previously 
reported work \cite{Patyal}, \emph{i.e.} $g$(powder) = 2.13 and $g$(c-axis) = 
2.148.  For the single crystal study, a single ESR line was observed at all 
temperatures, and the temperature dependence of the $g$ value is shown in Fig. 1.  
These data do not suggest the presence of a structural transition, and this result 
is consistent with neutron scattering studies \cite{Watson}.

\begin{figure}[hb]
\begin{center}\leavevmode
\includegraphics[width=0.8\linewidth]{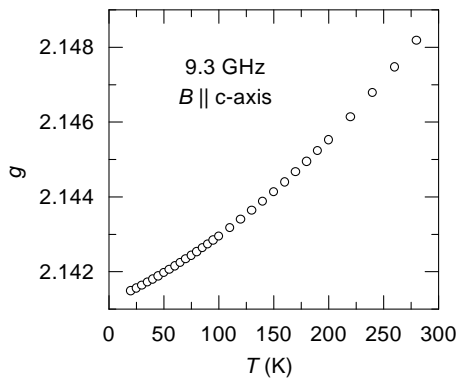}
\caption{The $g$ value of a 18.6 mg single 
crystal of BPCB as determined at 9.27 GHz and with $B\;||\:$c-axis.}
\label{Fig.1}\end{center}
\end{figure}

\begin{figure}[hb]
\begin{center}\leavevmode
\includegraphics[width=0.8\linewidth]{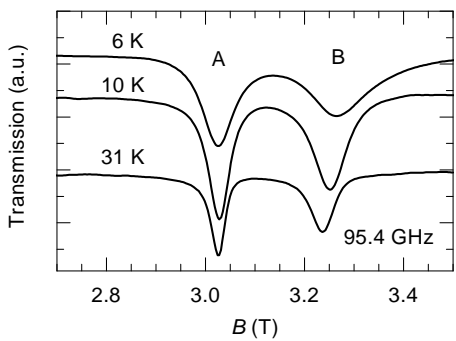}
\caption{Transmission, in arbitrary units, of BPCB at 95.4 GHz, 
where $B < B_{C1}$.}
\label{Fig.2}\end{center}
\end{figure}

Typical results obtained at 95.4 GHz are shown in Fig. 2, where two absorption 
lines are observed.  Assuming that these ESR spectra are predominantly probing 
the b-c$^*$ plane, then these results can be extrapolated to low fields, 
\emph{i.e.} 0.3 T, where the lines merge and reproduce the previously reported  
spectra \cite{Patyal}.  The integrated intensities of 
these absorption lines were obtained by using a Lorentzian lineshape to fit the 
data, and the temperature dependences of these intensities are shown in Fig. 3.  
Within the uncertainties, the temperature dependences of both lines are the 
same and similar to the results obtained from $\chi(T,B=0.1$ T).  
The data are consistent with a spin singlet ground state that is separated from 
the first excited triplet by an energy gap of $\approx 14$ K.  It is important 
to note that these spectra were collected below the first critical field, 
$B_{C1} = 6.6$ T \cite{Watson}.  It is noteworthy that one absorption line, A, 
corresponding to $g = 2.23$, does not change its position over the 
temperature range $3.7-42$ K, while line B shifts from $g=2.06$ at 3.7 K 
to $g= 2.09$ at 42 K.     

The ESR spectra at 289.7 GHz, where $B_{C1} < B < B_{C2}$
are shown in Fig. 4.  In this regime, the intensities of the lines 
increase with decreasing temperature, and this behavior is consistent 
with the magnetic field closing of the energy gap.  

A detailed interpretation of the high frequency results is difficult because  
the specimens were not single crystals.
Since Patyal \emph{et al.} \cite{Patyal} 
worked only at 300 K, a comparison to the temperature 
dependences of the $g$ values is not available.  
In summary, our results are consistent with earlier work 
\cite{Watson,Patyal}, and a complete analysis of the ESR spectra requires data from 
single crystals cooled to below the lowest exchange interaction, 
$J_{\parallel} = 3.8$ K. 

This work was supported, in part, by NSF DMR-0084173 
for the NHFML and the ACS-PRF 36163-AC5.  We have benefited from conversations with 
K.A. Abboud, J. \u{C}ern\'{a}k, and G.E. Granroth.

%
%

\begin{figure}[htb]
\begin{center}\leavevmode
\includegraphics[width=0.8\linewidth]{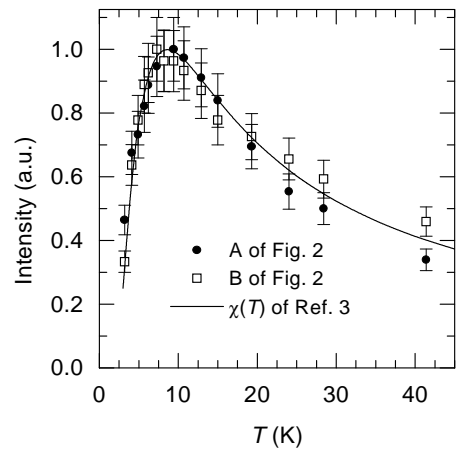}
\caption{The intensities, 
normalized to their largest values, in arbitrary units, 
of absorption lines A and B of Fig. 2.  
The solid line represents 
$\chi(T, B = 0.1$ T) \cite{Watson}.}
\label{Fig.3}\end{center}
\end{figure}

\begin{figure}[hb]
\begin{center}\leavevmode
\includegraphics[width=0.8\linewidth]{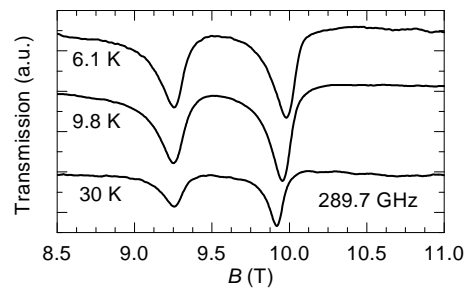}
\caption{Transmission, in arbitrary units, of BPCB at 289.7 GHz, 
where $B_{C1} < B < B_{C2}$.}
\label{Fig.4}\end{center}
\end{figure}

\end{document}